\documentclass[floatfix,twocolumn,aps,showpacs,showkeys,nofootinbib]{revtex4}

\usepackage{amssymb}
\usepackage{amsmath}
\usepackage{graphics}
\usepackage{epsfig}
\usepackage[dvips]{color}
\usepackage{bm}
\usepackage{color}

\newcommand{\be}{\begin{equation}}

\newcommand{\ee}{\end{equation}}
\newcommand{\bea}{\begin{eqnarray}}
\newcommand{\eea}{\end{eqnarray}}
\newcommand{\beas}{\begin{eqnarray*}}
\newcommand{\eeas}{\end{eqnarray*}}

\newcommand{\nn}{\nonumber\\}
\newcommand{\slsh}[1]{{\not \! #1}}
\newcommand{\slshh}[1]{{\not \!\! #1}}

\begin{document}
\title{Finite temperature quark-gluon vertex with a magnetic field in the Hard Thermal Loop approximation}
\author{Alejandro Ayala$^{1,5}$\footnote{Corresponding author: ayala@nucleares.unam.mx}, J.J. Cobos-Mart\1nez$^2$, M. Loewe$^{3,5}$, Mar\1a Elena Tejeda-Yeomans$^{4,1}$, R. Zamora$^3$}
\affiliation{$^1$Instituto de Ciencias
  Nucleares, Universidad Nacional Aut\'onoma de M\'exico, Apartado
  Postal 70-543, M\'exico Distrito Federal 04510,
  Mexico.\\
  $^2$Instituto de F\1sica y Matem\'aticas, Universidad Michoacana de San Nicol\'as de Hidalgo, Edificio C-3, Ciudad
  Universitaria, Morelia, Michoac\'an 58040, Mexico\\
  $^3$Instituto de F\1sica, Pontificia Universidad Cat\'olica de Chile,
  Casilla 306, Santiago 22, Chile.\\
  $^4$Departamento de F\1sica, Universidad de Sonora, Boulevard Luis Encinas J. y Rosales, Colonia Centro,
  Hermosillo, Sonora 83000, Mexico\\
  $^5$Centre for Theoretical and Mathematical Physics, and Department of Physics,
  University of Cape Town, Rondebosch 7700, South Africa}

\begin{abstract}

We compute the thermo-magnetic correction to the quark-gluon vertex in the presence of a weak magnetic field within the Hard Thermal Loop approximation. The vertex satisfies a QED-like Ward identity with the quark self-energy. Tihe only vertex components that get modified are the longitudinal ones. The calculation provides a first principles result for the quark anomalous magnetic moment at high temperature in a weak magnetic field.  We extract the effective thermo-magnetic quark-gluon coupling and show that this decreases as a function of the field strength. The result supports the idea that the properties of the effective quark-gluon coupling in the presence of a magnetic field are an important ingredient to understand the inverse magnetic catalysis phenomenon.

\end{abstract}

\pacs{11.10.Wx, 25.75.Nq, 98.62.En, 12.38.Cy}

\keywords{Chiral transition, Magnetic fields, Quark-gluon vertex, Quark anomalous magnetic moment.}

\maketitle

\section{Introduction}\label{Introduction}

The properties of strongly interacting matter under the influence of magnetic fields have been the subject of intense research over the last years. Lattice QCD results indicate that the transition temperature with 2+1 quark flavors decreases with increasing magnetic field strength~\cite{Fodor,Bali:2012zg,Bali2}. This behavior has been dubbed {\it inverse magnetic catalsis}. Possible explanations include a fermion paramagnetic contribution to the pressure with a sufficiently large magnetization~\cite{Noronha}, a back reaction of the Polyakov loop~\cite{Bruckmann:2013oba, Ferreira}, magnetic inhibition due to neutral meson fluctuations in a strong magnetic field~\cite{Fukushima}, high baryon density effects~\cite{Andersen2}, a decreasing magnetic field and temperature dependent coupling inspired by the QCD running of the coupling with energy, in the Nambu-Jona-Lasinio model~\cite{Farias, Ferreira1} and quark antiscreening from the effect of the anomalous magnetic moment of quarks in strong~\cite{Ferrer} and weak~\cite{Fayazbakhsh} magnetic fields. The phenomenon is not observed when only mean field approaches to describe the thermal environment are used~\cite{Andersen, Fraga, Loewe, Agasian, Mizher}, nor when calculations beyond mean field do not include magnetic effects on the coupling constants~\cite{ahmrv}.

In two recent studies~\cite{amlz,alz} we have shown that the decrease of the coupling constant with increasing field strength can be obtained within a perturbative calculation in the Abelian Higgs model and the Linear Sigma model, where charged fields are subject to the effect of a constant magnetic field. This behavior introduces a dependence of the boson masses on the magnetic field and a decrease of the critical temperature for chiral symmetry breaking/restoration. In order to establish whether a similar behavior takes place in QCD, the first step is the knowledge of the finite temperature and magnetic field dependence of the coupling constant. 

The purely magnetic field dependence of this coupling has been looked at in Ref.~\cite{Ferrer} for the case of a strong magnetic field. The authors find an anisotropic behavior and a decrease of the parallel part of the coupling with the field strength which in turn produces antiscreening and thus a decrease of the critical temperature for chiral symmetry restoration. However the strong field case is limited to extreme scenarios. For instance, in peripheral heavy ion collisions, though the initial intensity of the magnetic fields can be quite strong both at RHIC and the LHC, the field strength is a fast decreasing function of time~\cite{Kharzeev,Skokov,Bzdak}. By the time the gluons and quarks thermalize, the temperature becomes the largest of the energy scales. It seems that for most of the evolution of such systems a calculation that considers the strength of the magnetic field to be smaller than the square of the temperature is more appropriate.

The field theoretical treatment of systems involving massless bosons, such as gluons, at finite temperature in the presence of magnetic fields is plagued with subtleties. It is well known that unless a careful treatment is implemented, infrared divergences, associated to the effective dimensional reduction of the momentum integrals, appear. The divergence comes when accounting for the separation of the energy levels into transverse and longitudinal directions (with respect to the magnetic field direction). The former are given in terms of discrete Landau levels. Thus, the longitudinal mode alone no longer can tame the divergence of the Bose-Einstein distribution. 

This misbehavior can be overcome by a proper treatment of the physics involved when magnetic fields are introduced. For instance, it has recently been shown that it is possible to find the appropriate condensation conditions by accounting for the plasma screening effects~\cite{Ayala2}.  Here we show that a simple prescription where the fermion mass acts as the infrared regulator allows to obtain the leading behavior of the QCD coupling for weak magnetic fields at high temperature, that is, in the Hard Thermal Loop (HTL) approximation. As a first step, we compute the thermo-magnetic corrections to the quark-gluon vertex in the weak field approximation. We use this calculation to compute the thermo-magnetic dependence of the QCD coupling. To include the magnetic field effects we use Schwinger's proper time method. We should point out that the weak field approximation means that one considers the field strength to be smaller than the square of the temperature but does not imply a hierarchy with respect to other scales in the problem such as the fermion mass. 

The work is organized as follows: In Sec.~\ref{II} we recall how the magnetic field effects are included when charged virtual fermions are present. We set up the calculation of the Feynman diagrams involved. In Sec.~\ref{III} we work in the weak field approximation and find the thermo-magnetic behavior of the quark-gluon vertex. We show that the vertex thus found satisfies a QED-like Ward identity with the fermion self-energy computed within the same approximation. In Sec.~\ref{IV} we use this result to study the thermo-magnetic dependence of the coupling and show that this is a decreasing function of the magnetic field. We finally summarize and conclude in Sec.~\ref{conclusions}.

\section{Charged fermion propagator in a medium}\label{II}

The presence of a constant magnetic field breaks Lorentz invariance and leads to a charged fermion propagator which is function of the separate transverse and longitudinal momentum components (with respect to the field direction). Considering the case of a magnetic field pointing along the $\hat{z}$ direction, namely $\vec{B}=B\hat{z}$, the vector potential, in the so called {\it symmetric gauge}, is
\bea
   A_\mu(x)=\frac{B}{2}(0,-x_2,x_1,0).
\label{symgauge}
\eea
The fermion propagator in coordinate space cannot longer be written as a simple Fourier transform of a momentum propagator but instead it is written as~\cite{Schwinger}
\bea
   S(x,x')=\Phi (x,x')\int\frac{d^4p}{(2\pi)^4}e^{-ip\cdot (x-x')}S(k),
\label{genprop}
\eea
where
\bea
  \Phi (x,x')=\exp\left\{iq\int_{x'}^xd\xi^\mu\left[A_\mu + \frac{1}{2}F_{\mu\nu}(\xi - x')^\nu\right]\right\},
\label{phase}
\eea
is called the {\it phase factor} and $q$ is the absolute value of the fermion's charge, in units of the electron charge. $S(k)$ is given by
\bea
   S(k)&=&-i\int_0^\infty \frac{ds}{\cos (qBs)} e^{is(k_{\|}^2-k_\perp^2
   \frac{\tan (qBs)}{qBs} - m^2)}\nn
   &\times& \Big\{\left[\cos (qBs) + \gamma_1 \gamma_2 \sin (qBs) \right] (m+\slsh{k_{\|}}) \nn 
   &-& \frac{\slsh{k_\bot}}{\cos(qBs)} \Big\},
\label{Schwinger}
\eea
where $m$ is the quark mass and we use the definitions for the parallel and perpendicular components of the scalar product of two vectors $a^\mu$ and $b^\mu$ given by
\bea
   (a\cdot b)_{\|} &=& a_0b_0 - a_3b_3\nn
   (a\cdot b)_{\bot} &=& a_1b_1 + a_2b_2.
\label{defs}
\eea
\begin{widetext}
%%%%%%%%%%%%%%%%%%%%%%%%%%%%%%%%%%%
\begin{figure*}[t]
\begin{center}
\includegraphics[scale=0.45]{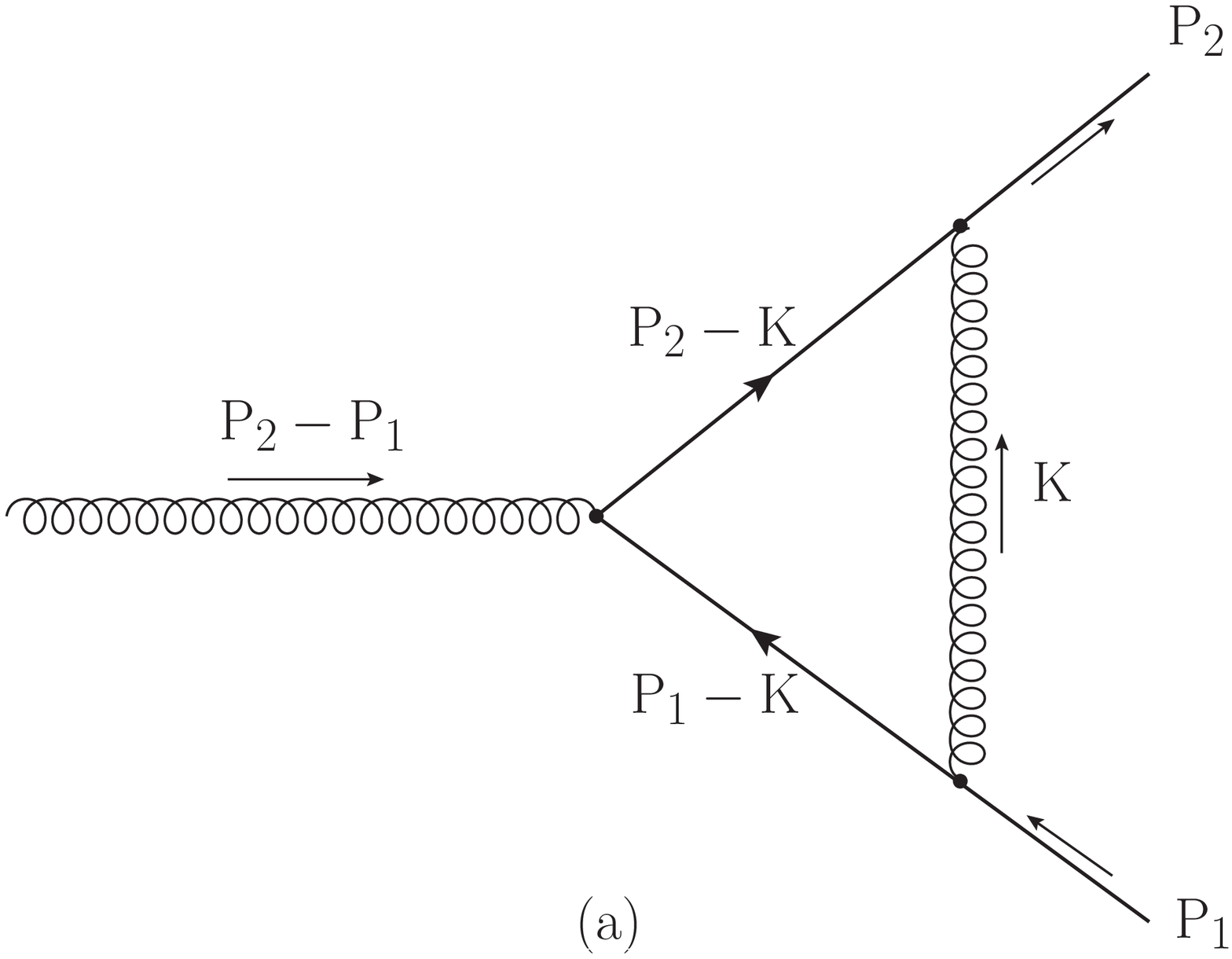}
\includegraphics[scale=0.45]{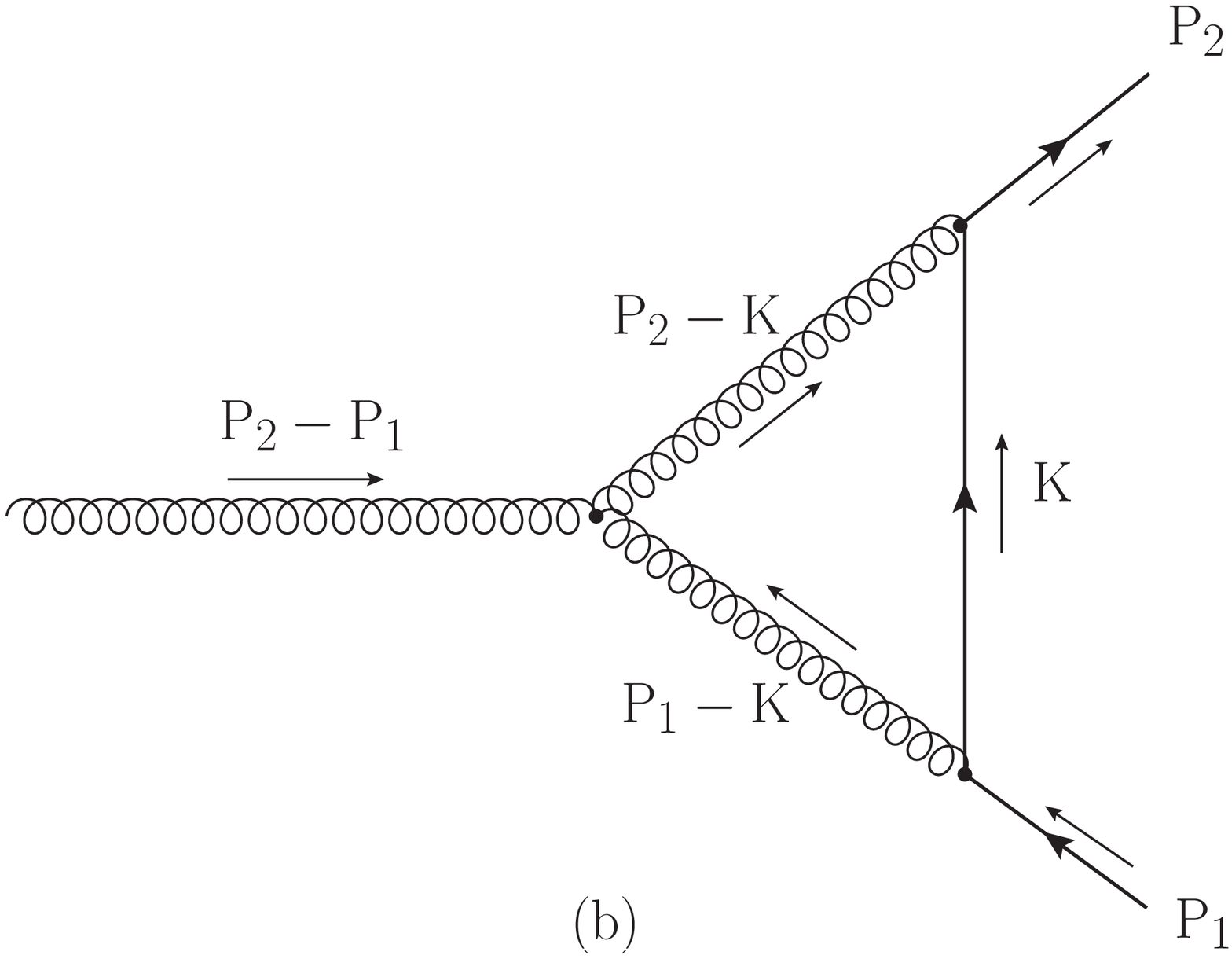}
\end{center}
\caption{Feynman diagrams contributing to the thermo-magnetic dependence of the quark-gluon vertex. Diagram (a) corresponds to a QED-like contribution whereas diagram (b) corresponds to a pure QCD contribution.}
\label{fig1}
\end{figure*}
%%%%%%%%%%%%%%%%%%%%%%%%%%%%%%%%%%%
\end{widetext}
Figure~\ref{fig1} shows the Feynman diagrams contributing to the quark-gluon vertex. Diagram (a) corresponds to a QED-like contribution whereas diagram (b) corresponds to a pure QCD contribution. The computation of these diagrams requires using the fermion propagator given by Eq.~(\ref{genprop}), which involves the phase factor in Eq.~(\ref{phase}).  Notice that the phase factor does not depend on the chosen path. Taking a straight line path parametrized by 
\be
   \xi^\mu=x'^\mu + t (x-x')^\mu,
\label{straight}
\ee
with $t \in [0,1]$, and using that $F_{\mu\nu}$ is antisymmetric, we can check that the phase factor becomes
\bea
   \Phi(x,x')=\exp\left\{\int_0^1dtA_\mu (x-x')^\mu\right\} ,
\label{newphase}
\eea
For $A_\mu$ given by Eq.~(\ref{symgauge}) the phase factor does not vanish. This factor can however be made to vanish by choosing an appropriate gauge transformation
\bea
   A_\mu (\xi)\rightarrow A'_\mu(\xi)=A_\mu + \frac{\partial}{\partial\xi^\mu} \Lambda (\xi),
\label{transf}
\eea
with
\bea
   \Lambda (\xi)&=&\frac{B}{2}(x'_2\xi_1-x'_1\xi_2)\nn
   \frac{\partial}{\partial\xi^\mu} \Lambda (\xi)&=&\frac{B}{2}(0,x'_2,-x'_1,0).
\label{lambda}
\eea
Therefore, when a single fermion propagator is involved, we can {\it gauge away} the phase factor and work with the momentum space representation of the propagator. The situation is similar when two fermion propagators stemming from a common vertex are involved. Choosing straight line paths, the product of the phase factors becomes
\bea
   \Phi (x,x')\Phi (x'',x)&=&\exp\left\{ iq\left(\int_{x'}^x\!\!d\xi^\mu A_\mu + \int_{x}^{x''}\!\!\!\!d\xi^\mu A_\mu
   \!\!\right)\right\}\nn
   &=& \exp\left\{ iq\int_{x'}^{x''}d\xi^\mu A_\mu\right\}\nn
   &=& \exp\left\{\int_0^1dtA_\mu (x''-x')^\mu\right\},
\label{twoprops}
\eea
where in the last step we have also chosen a straight line path connecting $x'$ and $x''$. We can now employ a gauge transformation similar to Eq.~(\ref{transf}) to gauge away the phase factor. Therefore, for the computation of diagrams (a) and (b) in Fig.~\ref{fig1}, we can just work with the momentum representation of the fermion propagators since the phase factors do not contribute. The situation would have been nontrivial in case the computation had required a three fermion propagator closed loop~\cite{Chyi}.

\section{QCD vertex at finite temperature with a weak magnetic field}\label{III}

To compute the leading magnetic field dependence of the vertex at high temperature, we work in the weak field limit of the momentum representation of the fermion propagator~\cite{Chyi}. We work in Euclidean space which is suited for calculations at finite temperature in the imaginary-time formalism of finite temperature field theory. The fermion propagator up to ${\mathcal{O}}(qB)$ is written as
\bea
   S(K)=\frac{m-\slshh{K}}{K^2+m^2} - i\gamma_1\gamma_2\frac{m-\slshh{K_{\|}}}{(K^2+m^2)^2}(qB).
\label{OqB}
\eea

Using the propagator in Eq.~(\ref{OqB}) and extracting a factor $gt_a$ common to the bare and purely thermal contributions to the vertex, the magnetic field dependent part of diagram (a) in Fig.~\ref{fig1} is expressed as
\bea
   \delta\Gamma_\mu^{\mbox{(a)}}&=&-ig^2(C_F - C_A/2)(qB)T\sum_n\int\frac{d^3k}{(2\pi)^3}\nn
   &\times&\gamma_\nu
   \left[ \gamma_1\gamma_2\slshh{K_{\|}}\gamma_\mu\slshh{K}\widetilde{\Delta}(P_2-K)\right.\nn
   &+& \left.
           \slshh{K}\gamma_\mu\gamma_1\gamma_2\slshh{K_{\|}}\widetilde{\Delta}(P_1-K)
   \right]\gamma_\nu\nn
   &\times&\Delta(K)\widetilde{\Delta}(P_2-K)\widetilde{\Delta}(P_1-K),
\label{(a)}
\eea
where in the spirit of the HTL approximation, we have ignored terms proportional to $m$ in the numerator and
\bea
   \widetilde{\Delta}(K)&\equiv&\frac{1}{\widetilde{\omega}_n^2 + k^2 + m^2}\nn
   \Delta(K)&\equiv&\frac{1}{\omega_n^2 + k^2}
\label{Deltas}
\eea
with $\widetilde{\omega}_n=(2n+1)\pi T$ and $\omega_n=2n\pi T$ the fermion and boson Matsubara frequencies. $C_F$, $C_A$ are the factors corresponding to the fundamental and adjoint representations of the $SU(N)$ Casimir operators
\bea
   C_F&=&\frac{N^2-1}{2N}\nn
   C_A&=&N,
\label{Casimir}
\eea 
respectively. Hereafter, capital letters are used to refer to four-momenta in Euclidean space with components $K_\mu=(k_4,\vec{k})=(-\omega,\vec{k})$, with $\omega$ either a Matsubara fermion or boson frequency. 

\begin{figure}[t]
\begin{center}
\includegraphics[scale=0.45]{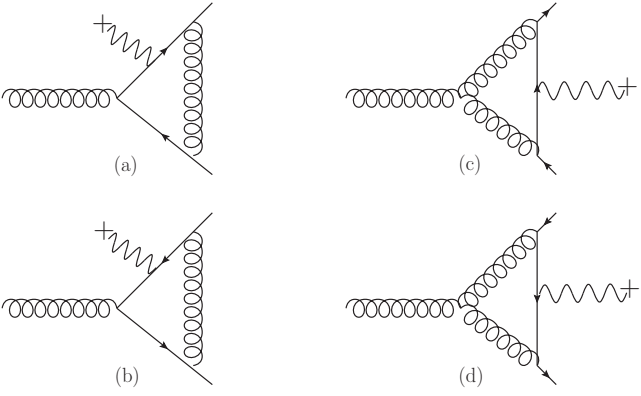}
\end{center}
\caption{Explicit Feynman diagrams accounting for the two fermion channels contributing to the quark-gluon vertex. The expansion of the fermion propagator at ${\mathcal{O}}(qB)$ can be represented by the insertion of a photon line attached from an external source to each internal fermion line.}
\label{fig2}
\end{figure}

In the same manner, the magnetic field dependent part of diagram (b) in Fig.~\ref{fig1} is expressed as
\bea
   \delta\Gamma_\mu^{\mbox{(b)}}&=&-2ig^2\frac{C_A}{2}(qB)T\sum_n\int\frac{d^3k}{(2\pi)^3}\nn
   &\times&\left[-\slshh{K}\gamma_1\gamma_2\slshh{K_{\|}}\gamma_\mu
                         +2\gamma_\nu\gamma_1\gamma_2\slshh{K_{\|}}\gamma_\nu K_\mu\right.\nn
   &-&\left.\gamma_\mu\gamma_1\gamma_2\slshh{K_{\|}}\slshh{K}
   \right]\nn
   &\times&\widetilde{\Delta}(K)^2\Delta (P_1-K)\Delta (P_2-K).
\label{(b)}
\eea
The explicit factor $2$ on the right-hand side of Eq.~(\ref{(b)}) accounts for the two possible fermion channels. These two channels are already accounted for in Eq.~(\ref{(a)}) since the magnetic field insertion on each quark internal line is thereby included at the order we are considering. The contribution from the two channels in each Feynman diagram is illustrated in Fig.~\ref{fig2}.

%%%%%%%%%%%%%%%%%%%%%%%%%%%%%%%%%%%
%%%%%%%%%%%%%%%%%%%%%%%%%%%%%%%%%%%
Recall that in Euclidean space it is convenient to work with the set of Dirac gamma matrices $\gamma_\mu$, $\mu=1,\ldots,4$, with
\bea
   \gamma_4=i\gamma_0,
\label{gamma4}
\eea
satisfying the algebra
\bea
   \left\{\gamma_\mu , \gamma_\nu\right\}=-2\delta_{\mu\nu}.
\label{algebra}
\eea
Also, $\gamma_5$ is defined by
\bea
   \gamma_5=\gamma_4\gamma_1\gamma_2\gamma_3,
\label{gama5}
\eea
which anticommutes with the rest of the gamma-matrices. 

We use this set of matrices and its properties to write
\bea
   \gamma_1\gamma_2\slshh{K_{\|}}=\gamma_5\left[(K\cdot b)\slsh{u} - (K\cdot u)\slsh{b}\right],
\label{rewrite}
\eea
where we have introduced the four-vectors
\bea
   u_\mu&=&(1,0,0,0)\nn
   b_\mu&=&(0,0,0,1),
\label{ub}
\eea
representing the medium's rest frame and the direction of the magnetic field, respectively. 

Note that in the HTL approximation $P_1$ and $P_2$ are small and can be taken to be of the same order. Thus, to extract the leading temperature behavior let us write 
\bea
   &&\left[ \gamma_1\gamma_2\slshh{K_{\|}}\gamma_\mu\slshh{K}\widetilde{\Delta}(P_2-K) +
           \slshh{K}\gamma_\mu\gamma_1\gamma_2\slshh{K_{\|}}\widetilde{\Delta}(P_1-K)
   \right]\nn
   &\simeq&
   \left[ \gamma_1\gamma_2\slshh{K_{\|}}\gamma_\mu\slshh{K} +
           \slshh{K}\gamma_\mu\gamma_1\gamma_2\slshh{K_{\|}}
   \right]\widetilde{\Delta}(P_1-K)
\label{HTL(a)}
\eea
in the integrand of Eq.~(\ref{(a)}). With the use of Eqs.~(\ref{rewrite}) and~(\ref{HTL(a)}) we proceed to work out the gamma-matrix structure in Eq.~(\ref{(a)}) and obtain
\bea
   \delta\Gamma_\mu^{\mbox{(a)}}&=&-2i\gamma_5g^2(C_F - C_A/2)(qB)\widetilde{G}_\mu(P_1,P_2),
\label{(a2)}
\eea
where
\bea
  \widetilde{G}_\mu(P_1,P_2)&=&2T\sum_n\int\frac{d^3k}{(2\pi)^3}\nn
  &\times& \Big\{
  (K\cdot b)\slshh{K}u_\mu - (K\cdot u)\slshh{K}b_\mu\nn
  &+& 
  \left[ (K\cdot b)\slsh{u} - (K\cdot u)\slsh{b}\right]K_\mu
  \Big\}\nn
  &\times&\Delta(K)\widetilde{\Delta}^2 (P_1-K)\widetilde{\Delta} (P_2-K).
\label{tildeGmu}
\eea
In the same fashion we work out the gamma-matrix structure in Eq.~(\ref{(b)}) and obtain
\bea
   \delta\Gamma_\mu^{\mbox{(b)}}&=&2i\gamma_5g^2\frac{C_A}{2}(qB)G_\mu(P_1,P_2),
\label{(b2)}
\eea
where
\bea
  G_\mu(P_1,P_2)&=&2T\sum_n\int\frac{d^3k}{(2\pi)^3}\nn
  &\times& \Big\{
  (K\cdot b)\slshh{K}u_\mu - (K\cdot u)\slshh{K}b_\mu\nn
  &+& 
  \left[ (K\cdot b)\slsh{u} - (K\cdot u)\slsh{b}\right]K_\mu
  \Big\}\nn
  &\times&\widetilde{\Delta}^2(K)\Delta(P_1-K)\Delta(P_2-K).
\label{Gmu}
\eea
Note that the gamma-matrix and vector structure of Eqs.~(\ref{tildeGmu}) and~(\ref{Gmu}) is the same. Also note that Eq.~(\ref{tildeGmu}) is obtained from Eq.~(\ref{Gmu}) by changing a boson line into a fermion line, which is accomplished by replacing the Bose-Einstein distribution $f(E)$ with minus a Fermi-Dirac distribution $-\widetilde{f}(E)$ and therefore this amounts for an overall change of sign~\cite{LeBellac}, namely
\bea
    \widetilde{G}_\mu (P_1,P_2)=-G_\mu (P_1,P_2).
\label{change}
\eea
Therefore, adding the contributions from the two Feynman diagrams in Fig.~\ref{fig1} we get
\bea
   \delta\Gamma_\mu&=&\delta\Gamma_\mu^{\mbox{(a)}}+\delta\Gamma_\mu^{\mbox{(b)}}\nn
   &=&2i\gamma_5g^2C_F(qB)G_\mu(P_1,P_2).
\label{deltaGamma}
\eea
$G_\mu (P_1,P_2)$ can be computed from the tensor ${\mathcal{J}}_{\alpha i}$ $(\alpha =1,\ldots 4,\ i=3,4)$ given by
\bea
   {\mathcal{J}}_{\alpha i}&=&T\sum_n\int\frac{d^3k}{(2\pi)^3}K_\alpha K_i\nn
   &\times&\widetilde{\Delta}^2(K)\Delta(P_1-K)\Delta(P_2-K),
   \label{mathcalJ}
\eea
which in turn requires to compute the frequency sums
\bea
   \widetilde{Y}_0&=&T\sum_n\widetilde{\Delta}^2(K)\Delta(P_1-K)\Delta(P_2-K)\nn
   &=&\left(-\frac{\partial}{\partial m^2}\right)T\sum_n\widetilde{\Delta}(K)\Delta(P_1-K)\Delta(P_2-K)\nn
   &\equiv&\left(-\frac{\partial}{\partial m^2}\right)\widetilde{X}_0\nn
   \widetilde{Y}_1&=&T\sum_n\omega_n\widetilde{\Delta}^2(K)\Delta(P_1-K)\Delta(P_2-K)\nn
   &=&\left(-\frac{\partial}{\partial m^2}\right)T\sum_n
   \omega_n\widetilde{\Delta}(K)\Delta(P_1-K)\Delta(P_2-K)\nn
   &\equiv&\left(-\frac{\partial}{\partial m^2}\right)\widetilde{X}_1\nn
   \widetilde{Y}_2&=&T\sum_n\omega_n^2\widetilde{\Delta}^2(K)\Delta(P_1-K)\Delta(P_2-K)\nn
   &=&\left(-\frac{\partial}{\partial m^2}\right)T\sum_n\omega_n^2\widetilde{\Delta}(K)\Delta(P_1-K)\Delta(P_2-K)\nn
   &\equiv&\left(-\frac{\partial}{\partial m^2}\right)\widetilde{X}_2,
\label{sums}
\eea
where $\widetilde{X}_0$, $\widetilde{X}_1$, $\widetilde{X}_2$  are in turn given by
\bea
   \widetilde{X}_0&=&-\sum_{s,s_1,s_2}\frac{ss_1s_2}{8EE_1E_2}
   \frac{1}{i(\omega_1-\omega_2)-s_1E_1+s_2E_2}\nn
   &\times&\left[\frac{1-\widetilde{f}(sE)+f(s_1E_1)}{i\omega_1-sE-s_1E_1} - 
   \frac{1-\widetilde{f}(sE)+f(s_2E_2)}{i\omega_2-sE-s_2E_2}
   	\right]\nn
	\widetilde{X}_1&=&i\sum_{s,s_1,s_2}\frac{s_1s_2E}{8EE_1E_2}
   \frac{1}{i(\omega_1-\omega_2)-s_1E_1+s_2E_2}\nn
   &\times&\left[\frac{1-\widetilde{f}(sE)+f(s_1E_1)}{i\omega_1-sE-s_1E_1} - 
   \frac{1-\widetilde{f}(sE)+f(s_2E_2)}{i\omega_2-sE-s_2E_2}
   	\right]\nn
   \widetilde{X}_2&=&\sum_{s,s_1,s_2}\frac{ss_1s_2E^2}{8EE_1E_2}
   \frac{1}{i(\omega_1-\omega_2)-s_1E_1+s_2E_2}\nn
   &\times&\left[\frac{1-\widetilde{f}(sE)+f(s_1E_1)}{i\omega_1-sE-s_1E_1} - 
   \frac{1-\widetilde{f}(sE)+f(s_2E_2)}{i\omega_2-sE-s_2E_2}
   	\right].\nn
\label{tildeXs}
\eea

The leading temperature behavior is obtained from the terms with $s=-s_1=-s_2$. Let us consider in detail the calculation of $\widetilde{X}_0$ for those terms. We make the approximation where $f(E_1)\simeq f(E_2)\simeq f(E)$, namely, that the Bose-Einstein distribution depends of $E=\sqrt{k^2+m^2}$ and thus on the quark mass. This approximation allows to find the leading temperature behavior for $m\rightarrow 0$ since it amounts to keep the quark mass as an infrared regulator. Also, using that $E_i\simeq k - \vec{p}_i\cdot\hat{k}$, $i=1,2$, we get  
\bea
   \widetilde{X}_0&\simeq&-\frac{1}{8k^2}\frac{\left[\widetilde{f}(E)+f(E)\right]}{E}\nn
   &\times&\left\{
   \frac{1}{(i\omega_1 + \vec{p}_1\cdot\hat{k})(i\omega_2 + \vec{p}_2\cdot\hat{k})} \right.\nn
   &+& \left.
   \frac{1}{(i\omega_1 - \vec{p}_1\cdot\hat{k})(i\omega_2 - \vec{p}_2\cdot\hat{k})}
   \right\},
\label{HTLX0}
\eea
where in the denominator of the first fraction we have set $E_1=E_2=k$. In a similar fashion
\bea
   \widetilde{X}_1&\simeq&-\frac{i}{8k}\frac{\left[\widetilde{f}(E)+f(E)\right]}{E}\nn
   &\times&\left\{
   \frac{1}{(i\omega_1 + \vec{p}_1\cdot\hat{k})(i\omega_2 + \vec{p}_2\cdot\hat{k})}\right. \nn
   &-& \left.
   \frac{1}{(i\omega_1 - \vec{p}_1\cdot\hat{k})(i\omega_2 - \vec{p}_2\cdot\hat{k})}
   \right\}\nn
   \widetilde{X}_2&\simeq&\frac{1}{8}\frac{\left[\widetilde{f}(E)+f(E)\right]}{E}\nn
   &\times&\left\{
   \frac{1}{(i\omega_1 + \vec{p}_1\cdot\hat{k})(i\omega_2 + \vec{p}_2\cdot\hat{k})} \right.\nn
   &+& \left.
   \frac{1}{(i\omega_1 - \vec{p}_1\cdot\hat{k})(i\omega_2 - \vec{p}_2\cdot\hat{k})}
   \right\}.
\label{HTLX1}
\eea

Using Eqs.~(\ref{HTLX0}) and~(\ref{HTLX1}) into Eqs.~(\ref{sums}) and~(\ref{mathcalJ}), we get
\bea
   {\mathcal{J}}_{\alpha i}&=&-\frac{1}{8\pi^2}\left(-\frac{\partial}{\partial y^2}\right)\int_0^\infty 
   \frac{dx\ x^2}{\sqrt{x^2+y^2}}\nn
   &\times&\left[\widetilde{f}(\sqrt{x^2+y^2})+f(\sqrt{x^2+y^2})\right]\nn
   &\times&\int\frac{d\Omega}{4\pi}
   \frac{\hat{K}_\alpha \hat{K}_i}{(P_1\cdot\hat{K})(P_2\cdot\hat{K})},
\label{integr}
\eea
where we defined $x=k/T$, $y=m/T$, $\hat{K}=(-i,\hat{k})$, $P_1=(-\omega_1,\vec{p}_1)$ and $P_2=(-\omega_2,\vec{p}_2)$. The integrals over $x$ can be expressed in terms of the well known functions~\cite{Jackiw}
\bea
   h_n(y)&=&\frac{1}{\Gamma (n)}\int_0^\infty 
   \frac{dx\ x^{n-1}}{\sqrt{x^2+y^2}}\frac{1}{e^{\sqrt{x^2+y^2}}-1}\nn
   f_n(y)&=&\frac{1}{\Gamma (n)}\int_0^\infty 
   \frac{dx\ x^{n-1}}{\sqrt{x^2+y^2}}\frac{1}{e^{\sqrt{x^2+y^2}}+1},
\label{handf}
\eea
which satisfy the differential equations
\bea
   \frac{\partial h_{n+1}}{\partial y^2}&=&-\frac{h_{n-1}}{2n}\nn
   \frac{\partial f_{n+1}}{\partial y^2}&=&-\frac{f_{n-1}}{2n},
\label{diffeq}
\eea
therefore
\bea
   {\mathcal{J}}_{\alpha i}&=&-\frac{1}{16\pi^2}\left[h_1(y)+f_1(y)\right]\nn
   &\times&\int\frac{d\Omega}{4\pi}
   \frac{\hat{K}_\alpha \hat{K}_i}{(P_1\cdot\hat{K})(P_2\cdot\hat{K})}.
\label{simpl}
\eea
Using the high temperature expansions for $h_1(y)$ and $f_1(y)$~\cite{Kapusta}
\bea
   h_1(y)&=&\frac{\pi}{2y} + \frac{1}{2}\ln\left(\frac{y}{4\pi}\right) + \frac{1}{2}\gamma_E + \ldots\nn
   f_1(y)&=&-\frac{1}{2}\ln\left(\frac{y}{\pi}\right) - \frac{1}{2}\gamma_E + \ldots,
\label{expansions}
\eea
and keeping the leading terms, we get
\bea
   {\mathcal{J}}_{\alpha i}=\frac{1}{16\pi^2}\left[\ln(2) - \frac{\pi}{2}\frac{T}{m}\right]
   \int\frac{d\Omega}{4\pi}
   \frac{\hat{K}_\alpha \hat{K}_i}{(P_1\cdot\hat{K})(P_2\cdot\hat{K})}.
\label{finally}
\eea
Using Eqs.~(\ref{finally}) and~(\ref{Gmu}) into Eq.~(\ref{deltaGamma}) we obtain
\bea
  \delta\Gamma_\mu(P_1,P_2)&=&4i\gamma_5g^2C_FM^2(T,m,qB)\nn
  &\times& \int\frac{d\Omega}{4\pi}\frac{1}{(P_1\cdot\hat{K})(P_2\cdot\hat{K})}\nn
  &\times&
  \Big\{
  (\hat{K}\cdot b)\slshh{\hat{K}}u_\mu - (\hat{K}\cdot u)\slshh{\hat{K}}b_\mu\nn
  &+& 
  \left[ (\hat{K}\cdot b)\slsh{u} - (\hat{K}\cdot u)\slsh{b}\right]\hat{K}_\mu
  \Big\}.
\label{Gmuexpl}
\eea
where we have defined de function $M^2(T,m,qB)$ as
\bea
   M^2(T,m,qB)=\frac{qB}{16\pi^2}\left[\ln(2) - \frac{\pi}{2}\frac{T}{m}\right].
\label{M}
\eea
%%%%%%%%%%%%%%%%%%%%%%%%%%%%%%%%%%%
\begin{figure}[t]
\begin{center}
\includegraphics[scale=0.45]{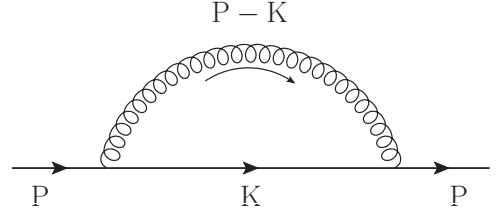}
\end{center}
\caption{Feynman diagram for the the quark self-energy. The internal quark line represents the quark propagator in the presence of the magnetic field in the weak field limit.}
\label{fig3}
\end{figure}
%%%%%%%%%%%%%%%%%%%%%%%%%%%%%%%%%%%

It is no surprise that the thermo-magnetic correction to the quark-gluon vertex is proportional to $\gamma_5$ since the magnetic field is odd under parity conjugation. Furthermore, it is important to note that the vertex thus found satisfies a QED-like Ward identity. The simplest way to see this is to look at Eq.~(\ref{deltaGamma}) with $G_\mu (P_1,P_2)$ given by Eq.~(\ref{Gmu}). Contracting this function with $(P_1-P_2)_\mu$ we get
\bea
  (P_1-P_2)\cdot G(P_1,P_2)&=&2T\sum_n\int\frac{d^3k}{(2\pi)^3}\nn
  &\times& \Big\{
  (K\cdot b)\slshh{K}(P_1-P_2)\cdot u\nn
  &-& (K\cdot u)\slshh{K}(P_1-P_2)\cdot b\nn
  &+& 
  \left[ (K\cdot b)\slsh{u} - (K\cdot u)\slsh{b}\right]\nn
  &\times&(P_1-P_2)\cdot K
  \Big\}\widetilde{\Delta}^2(K)\nn
  &\times&\Delta(P_1-K)\Delta(P_2-K).\nn
\label{W1}
\eea
In the spirit of the HTL approximation we ignore loose factors of $P_1$, $P_2$ in the numerator. Noting that
\bea
   (P_1-P_2)\cdot K&\simeq& \frac{1}{2}(P_2-K)^2 - \frac{1}{2}(P_1-K)^2\nn
   &=&\frac{1}{2}\Delta^{-1}(P_2-K) - \frac{1}{2}\Delta^{-1}(P_1-K),\nn
\label{approxinHTL}
\eea
we get
\bea
  (P_1-P_2)\cdot G(P_1,P_2)&=&T\sum_n\int\frac{d^3k}{(2\pi)^3}\nn
  &\times&
  \Big\{ (K\cdot b)\slsh{u} - (K\cdot u)\slsh{b}\Big\}\nn
  &\times&\left[\Delta(P_1-K)\widetilde{\Delta}^2(K)\right.\nn
  &-& \left.\Delta(P_2-K)\widetilde{\Delta}^2(K)\right].
\label{W2}
\eea
Therefore, the expression in Eq.~(\ref{deltaGamma}) satisfies the QED-like Ward identity, in the HTL approximation, given by
\bea
   (P_1-P_2)\cdot\delta\Gamma (P_1,P_2)= \Sigma (P_1) - \Sigma (P_2),
\label{W2}
\eea
where, as can be computed from the diagram in Fig.~\ref{fig3}, the quark self-energy in the presence of a weak magnetic field in the HTL approximation is given by
\bea
    \Sigma (P) &=& 2i\gamma_5g^2C_F(qB)
    T\sum_n\int\frac{d^3k}{(2\pi)^3}\Delta(P-K)\widetilde{\Delta}^2(K)\nn
  &\times&
  \Big\{ (K\cdot b)\slsh{u} - (K\cdot u)\slsh{b}\Big\}.
\label{selfenergy}
\eea
Using the same approach that lead to finding the expression for the vertex, we get the explicit expression for the self-energy, given by
\bea
   \Sigma(P)&=&2i\gamma_5g^2C_FM^2(T,m,qB)\nn
  &\times& \int\frac{d\Omega}{4\pi}\frac{\left[ (\hat{K}\cdot b)\slsh{u} - (\hat{K}\cdot u)\slsh{b}\right]}
  {(P\cdot\hat{K})}.
\label{selfexpl}
\eea
This remarkable result shows that even in the presence of the magnetic field and provided the temperature is the largest of the energy scales, the thermo-magnetic correction to the quark-gluon vertex is gauge invariant.

\section{Thermo-magnetic QCD coupling}\label{IV}

In order to look at the thermo-magnetic dependence of the quark-gluon coupling let us look explicitly at the quantities
\bea
   J_{\alpha i}(P_1,P_2)\equiv\int\frac{d\Omega}{4\pi}\frac{\hat{K}_\alpha\hat{K}_i}
   {(P_1\cdot{\hat{K}})(P_2\cdot{\hat{K}})},
\label{Js}
\eea
appearing on the r.h.s. of Eq.~(\ref{Gmuexpl}). For the sake of simplicity, let us choose a configuration where the momenta $\vec{p}_1$ and $\vec{p}_2$ make a relative angle $\theta_{12}=\pi$. This configuration corresponds for instance to a thermal gluon decaying into a quark-antiquark pair in the center of mass system and is therefore general enough. Consider first $J_{44}(P_1,P_2)$
\bea
   J_{44}(P_1,P_2)&=&-\frac{1}{2}\frac{1}{i\omega_1p_2+i\omega_2p_1}\nn
   &\times&
   \int_{-1}^1dx\left\{\frac{p_1}{i\omega_1 + p_1x} + \frac{p_2}{i\omega_2 - p_2x}\right\}\nn
   &=&-\frac{1}{2}\frac{1}{i\omega_1p_2+i\omega_2p_1}\nn
   &\times&
   \left\{ \ln\left(\frac{i\omega_1 + p_1}{i\omega_1 - p_1}\right) + \ln\left(\frac{i\omega_2 + p_2}{i\omega_2 - p_2}\right)
   \right\}.\nn
\label{J44}
\eea
We now take the analytic continuation to Minkowski space $i\omega_{1,2}\rightarrow p_{0 1,0 2}$ [$\hat{K}\rightarrow (-1,\hat{k})$] and consider the scenario where $p_{0 1}=p_{0 2}\equiv p_0$ and $p_1=p_2\equiv p$, then we get
\bea
   J_{44}\rightarrow J_{00}=\frac{1}{2p_0p}\ln\left(\frac{p_0+p}{p_0-p}\right).
\label{J44toMink}
\eea
Furthermore, let us look now at the {\it static limit}, that is, where the quarks are almost at rest, namely $p\rightarrow 0$, then 
\bea
   J_{00}\stackrel{p\rightarrow 0}{\longrightarrow}\frac{1}{p_0^2}.
\label{J00inthelim}
\eea
Now consider $J_{33}(P_1,P_2)$ in the same momenta  configuration
\bea
   J_{33}(P_1,P_2)&=&\frac{1}{2}\frac{1}{i\omega_1p_2+i\omega_2p_1}\nn
   &\times&
   \int_{-1}^1dx\ x^2\left\{\frac{p_1}{i\omega_1 + p_1x} + \frac{p_2}{i\omega_2 - p_2x}\right\}\nn
   &=&-\frac{1}{i\omega_1p_2+i\omega_2p_1}\nn
   &\times& \left\{
   \frac{i\omega_1}{p_1}\left[ 1 - \frac{i\omega_1}{2p_1}\ln\left(\frac{i\omega_1 + p_1}{i\omega_1 - p_1}\right)\right]
   \right.\nn
   &+& \left. \frac{i\omega_2}{p_2}\left[ 1 - \frac{i\omega_2}{2p_2}\ln\left(\frac{i\omega_2 + p_2}{i\omega_2 - p_2}\right)\right]
   \right\}.\nn
\label{J33inic}
\eea
After analytical continuation to Minkowski space and in the same scenario where $p_{0 1}=p_{0 2}\equiv p_0$ and $p_1=p_2\equiv p$, we get
\bea
   J_{33}=-\frac{1}{p^2}\left[1 - \frac{p_0}{2p}\ln\left(\frac{p_0+p}{p_0-p}\right)\right].
\label{J33after}
\eea
In the limit where $p\rightarrow 0$,
\bea
    J_{33}\stackrel{p\rightarrow 0}{\longrightarrow}\frac{1}{3p_0^2}.
\label{J33}
\eea
For the same choice of momenta, the rest of the components of $J_{\alpha i}$ vanish, which means that only the longitudinal components of the thermo-magnetic vertex are modified. Using Eqs.~(\ref{Js}),~(\ref{J44toMink}) and~(\ref{J33}) into Eq.~(\ref{Gmuexpl}), the explicit longitudinal components are given by
\bea
  \vec{\delta\Gamma}_\parallel(p_0) = \left(\frac{2}{3p_0^2}\right)4g^2C_FM^2(T,m,qB)\vec{\gamma}_\parallel\Sigma_3,
\label{components}
\eea
where $\vec{\gamma}_\parallel = (\gamma_0,0,0,-\gamma_3)$ and we have rearranged the gamma-matrices to introduce the spin operator in the $\hat{z}$-direction
\bea
   \Sigma_3&=&i\gamma_1\gamma_2\nn
   &=&\frac{i}{2}[\gamma_1,\gamma_2].
\label{spinop}
\eea 

That  the vertex correction in Eq.~(\ref{components}) is proportional to the third component of the spin operator is of course natural since the first order magnetic correction to the vertex is in turn proportional to the spin interaction with the magnetic field, which we chose to point along the third spatial direction. The correction thus corresponds to the quark anomalous magnetic moment at high temperature in a weak magnetic field. Note also that Eq.~(\ref{components}) depends on the scales $p_0$ and $m$. $p_0$ is the typical energy of a quark in the medium and therefore the simplest choice for this scale is to take it as the temperature. The quark mass represents the infrared scale and it is therefore natural to take it as the thermal quark mass. We thus set
\bea
   p_0&=&T\nn
   m^2&=&m_f^2=\frac{1}{8}g^2T^2C_F.
\label{choise}
\eea

\begin{figure}[t]
\begin{center}
\includegraphics[scale=0.55]{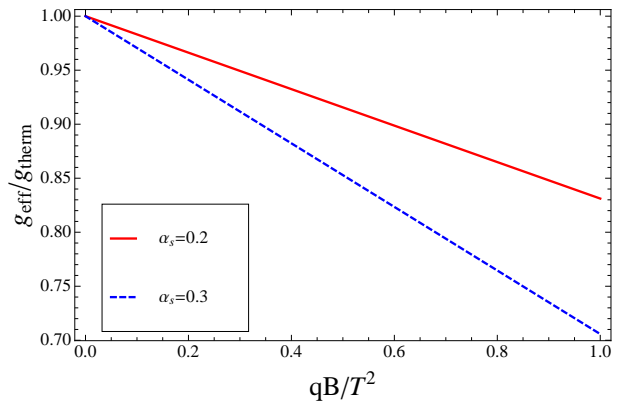}
\end{center}
\caption{The effective thermo-magnetic coupling $g_{\mbox{\small{eff}}}$ normalized to the purely thermal coupling $g_{\mbox{\small{therm}}}$ as a function of the field strength scaled by the squared of the temperature, for $\alpha_s=0.2, 0.3$. Note that $g_{\mbox{\small{eff}}}$ decreases down to about 15\% -- 25\% for the largest strength of the magnetic field within the weak field limit with respect to the purely thermal correction and that the decrease is faster for larger $\alpha_s$.}
\label{fig4}
\end{figure}

As is well known, the purely thermal correction to the quark-gluon vertex is given by~\cite{LeBellac}
\bea
   \delta\Gamma^{\mbox{\small{therm}}}_\mu(P_1,P_2)=-m_f^2\int\frac{d\Omega}{4\pi}
   \frac{\hat{K}_\mu\slshh{\hat{K}}}{(P_1\cdot\hat{K})(P_2\cdot\hat{K})}.
\label{purether}
\eea
In order to extract the effective modification to the coupling constant in one of the longitudinal directions, let us look at the contribution proportional to $\gamma_0$ from Eq.~(\ref{purether}). For the same working momenta configuration and using Eqs.~(\ref{J44toMink}) and~(\ref{J00inthelim})
\bea
   \delta\Gamma^{\mbox{\small{therm}}}_0(p_0)=-\frac{m_f^2}{p_0^2}\gamma_0.
\label{purether0}
\eea
Using Eqs.~(\ref{components}) and~(\ref{purether0}), the effective thermo-magnetic modification to the quark-gluon coupling extracted from the effective longitudinal vertex, is given by
\bea
   g_{\mbox{\small{eff}}}=g\left[1-\frac{m_f^2}{T^2}+\left(\frac{8}{3T^2}\right)g^2C_FM^2(T,m_f,qB)\right],
\label{effective}
\eea
where we have used a spin configuration with eigenvalue 1 for $\Sigma_3$. Figure~\ref{fig4} shows the behavior of $g_{\mbox{\small{eff}}}$ normalized to $g_{\mbox{\small{therm}}}$, where
\bea
   g_{\mbox{\small{therm}}}\equiv g\left(1-\frac{m_f^2}{T^2}\right),
\label{effectivetherm}
\eea
for $\alpha_s=g^2/4\pi =0.2, 0.3$ as a function of the scaled variable $b=qB/T^2$. Plotted as a function of $b$ and for our choice of momentum scales, the function $g_{\mbox{\small{eff}}}/g_{\mbox{\small{therm}}}$ is temperature independent. Note that the effective thermo-magnetic coupling $g_{\mbox{\small{eff}}}$ decreases as a function of the magnetic field. The decrease becomes more significant for larger $\alpha_s$ and for the considered values of $\alpha_s$ it becomes about 15\% -- 25\% smaller than the purely thermal correction for $qB\sim T^2\sim 1$.

%%%%%%%%%%%%%%%%%%%%%%%%%%%%%%%%%%%
%%%%%%%%%%%%%%%%%%%%%%%%%%%%%%%%%%%

\section{Summary and Conclusions}\label{conclusions}

In this work we have computed the thermo-magnetic corrections to the quark-gluon vertex for a weak magnetic field and in the HTL approximation. We have shown that this vertex satisfies a QED-like Ward identity with the quark self-energy. This results hints to the gauge-independence of the calculation. The thermo-magnetic correction is proportional to the spin component in the direction of the magnetic field and affects only the longitudinal components of the quark-gluon vertex. It thus corresponds to the quark anomalous magnetic moment at high temperature in a weak magnetic field. 

From the quark-gluon vertex we have extracted the behavior of the magnetic field dependence of the QCD coupling. We have shown that the coupling decreases as the field strength increases. For analytic simplicity, the explicit calculation has been performed for a momentum configuration that accounts for the conditions prevailing in a quark-gluon medium at high temperature, namely namely back-to-back slow moving quarks whose energy is of the order of the temperature and with the infrared scale of the order of the thermal particle's mass. The chosen values for these scales provide a good representation of the coupling constant's strength within the plasma conditions.

We stress that the weak field approximation means that the field strength is smaller than the square of the temperature but does not require a hierarchy with respect to other scales in the problem such as the fermion thermal mass. 

The result supports the idea~\cite{amlz, alz} that the decreasing of the coupling constant is an important ingredient to understand the inverse magnetic catalysis obtained in lattice QCD. It remains to include this result within a calculation to allow the extraction of the magnetic field dependence of the critical temperature for chiral symmetry restoration/deconfinement transition in QCD. This is work in progress and will be reported elsewhere.

\section*{Acknowledgments}

A. A. acknowledges useful conversations with G. Krein. Support for this work has been received in part from CONACyT-M\'exico under grant number 128534 and FONDECYT under grant  numbers 1130056 and 1120770. R. Z. acknowledges support from CONICYT under Grant No. 21110295.

\end{document}